\documentclass[a4paper,11pt]{article}
\pdfoutput=1 

\usepackage{jheppub} 
\usepackage{multirow}

\usepackage[T1]{fontenc} 

\title{\boldmath Gaussian null coordinates, near-horizon geometry and conserved charges on the horizon of extremal non-dilatonic black $p$-branes\footnote{Based in part on a lecture given by M.C. at Fifth Amazonian Symposium on Physics, Bel\'{e}m, November 18-22, 2019.}}

\author[a,b]{Mirjam Cveti{\v{c}},}
\author[a,c]{Paulo J. Porf\'{\i}rio,}
\author[d]{Alejandro Satz}

\affiliation[a]{Department of Physics and Astronomy, University of Pennsylvania,
Philadelphia, PA 19104, USA}
\affiliation[b]{Center for Applied Mathematics and Theoretical Physics, University of Maribor, SI2000 Maribor, Slovenia}
\affiliation[c]{Departamento de F\'{\i}sica, Universidade Federal da Para\'{\i}ba, Caixa Postal 5008, 58051-970, Jo\~ao Pessoa, Para\'{\i}ba, Brazil}
\affiliation[d]{Sarah Lawrence College, Bronxville, NY 10708, USA}
\emailAdd{cvetic@physics.upenn.edu}
\emailAdd{fepa@sas.upenn.edu}
\emailAdd{asatz@sarahlawrence.edu}

\abstract{In this paper, we examine the emergence of conserved charges on the horizon of a particular class of extremal non-dilatonic  black $p$-branes (which reduce to extremal dilatonic black holes in $D=4$ dimensions upon toroidal compactification) in the presence of a probe massless scalar field in the bulk. This result is achieved by writing the black $p$-brane geometry in a Gaussian null coordinate system which allows us to get a non-singular near-horizon geometry description. We find that the near-horizon geometry is $AdS_{p+2}\times S^2$ and that the $AdS_{p+2}$ section has an internal structure which can be seen as a warped product of $AdS_{2}\times S^{p}$ in Gaussian null coordinates. We show that the bulk scalar field satisfying the field equations is expanded in terms of non-normalizable and normalizable modes, which for certain suitable quantization conditions are well-behaved at the boundary of $AdS_{p+2}$ space. Furthermore, we show that picking the normalizable modes results in the existence of conserved quantities on the horizon. We discuss  the impact of these conserved quantities in the late time regime.  
}

\begin{document}
\maketitle
\flushbottom

\section{Introduction}

The search for a full theory of quantum gravity has  intensified over the last decades, with string theory as one of the most prominent candidates. In this scenario, black holes and extended objects as $p/D$-branes play an important role (see f.e. \cite{Stelle:1996tz, Mohaupt:2000gc, Horowitz:1991cd, Skenderis:1999bs, Duff:1994an} for a review). These objects display remarkable properties, for example, black holes from General Relativity (GR) are thermodynamical systems  characterized by a temperature and an entropy which is proportional to the area of the horizon. It is expected that the microscopic origin of the entropy of black holes should be explained by a fully quantum gravity theory. In the context of extended objects, the first law of black branes mechanics has been discussed in \cite{Townsend:2001rg}.

A particular wave of interest has  risen recently around extremal black holes and the  near-extremality condition.  In particular, Aretakis has investigated the stability of extremal black holes \cite{Aretakis:2011gz, Aretakis:2011ha, Aretakis:2011hc, Aretakis:2012ei} by perturbing the spacetime geometry with probe fluctuating fields. An especially interesting result from this study is the existence of a  conserved charge on the horizon called Aretakis charge or constant which itself sets up as a ``hair". This effect is remarkable because the radial derivative of a massless scalar field ($\partial_{r}\psi$) decays outside the horizon  while in the horizon it becomes a constant. Aretakis also showed that any higher radial derivative ($\partial_{r}^{k}\psi$) blows up at the horizon. The Aretakis charge was recently calculated \cite{Cvetic:2020zqb} for extremal rotating STU black holes in four-dimensions \cite{Cvetic:1996xz} and five-dimensions \cite{Cvetic:1996kv}.

It has been shown in \cite{Godazgar:2017igz, Cvetic:2018gss} that the Aretakis charge for extremal Reissner-Nordstrom (RN) black holes perturbed by a massless scalar field is intimately related to conserved charges at the future null infinity, called Newman-Penrose charges \cite{Newman:1968uj}, which are present in any asymptotically flat spacetime. The duality between both charges is related to a conformal symmetry linking the RN metric to its  transformation under inversion of the radial coordinate, which converts the horizon into  future null infinity. Aside from this, the massless wave equation displays an additional conformal symmetry on this background, called Couch-Torrence symmetry \cite{couch}, which allows a direct mapping between the scalar field near the horizon and the scalar field at the future null infinity say. Consequently, it is possible to map  Aretakis charges to  NP charges.  

Most 
papers have explored these subjects in the context of four-dimensional black holes; however, in \cite{Figueras:2008qh} the authors have treated such issues in higher-dimensional black holes. On the other hand, the approach used there does not apply in the context of extended objects as black $p$-branes since it requires the assumption of the horizon should be compact, which is not the case for higher-dimensional branes. Hence, a question should be raised: Are Aretakis charges conserved on non-compact horizons? This study aims at to addressing this question, which has not been explored in the literature. Our main goal is  to examine the existence of conserved charges on the horizon of extremal non-dilatonic black branes. At this moment, it is worth calling attention for some aspects in this scenario. The first one is that there is no general Gaussian null coordinates prescription to describe extended objects in a distinguishing way from black holes. In fact, such a coordinate system becomes crucial in order to gather information near the horizon.
Secondly, it is well known that a large set of $p/D$-branes interpolates between $AdS_{p+2}\times S^{D-(p+2)}$ geometry in the near-horizon limit and  $D$-dimensional Minkowski space in the asymptotic limit \cite{Gibbons:1993sv}.

It has been shown in \cite{Gibbons:1994vm} that extremal dilatonic black holes can be uplifted to extremal non-dilatonic black $p$-branes. This relation between both allows us to resolve in many cases of interest the problem of the metric and dilaton blowing up at the Killing horizon  by just uplifting the Einstein-Maxwell-dilaton theory in four dimensions to 
the Einstein-Maxwell one in higher dimensions. 
This was first explored in \cite{Gibbons:1993sv}, where the authors have  considered the asymptotic behavior of the metric and dilaton near the horizon. As aforementioned, the asymptotic metric of the extremal black $p$-brane near the horizon is just a product of a lower-dimensional anti de-Sitter space with a sphere, and upon double-dimensional reduction it reduces to a conformally $AdS_2\times S^2$ metric of the Einstein-Maxwell-dilaton theory, describing the near-horizon of extremal dilatonic black holes with the dilaton coupling being $a=\sqrt{\dfrac{p}{p+2}}$. Furthermore, the holographic description of such near-horizon geometries upon Kaluza-Klein reduction on the compact space
 was laid out in  \cite{Maldacena:1997re,Witten:1998qj,Klebanov:1999tb}.

Our goal in this work will be to provide a new avenue for testing black $p$-branes instabilities.  We  start our analysis by perturbing an extremal non-dilatonic black $p$-brane in the near-horizon limit with a massless scalar field. Such a scalar field is dealt with as a probe that does not have enough energy to warp the space-time geometry. Therefore, the problem reduces to solving the massless Klein-Gordon equation in the near-horizon background. We will proceed with a coordinate transformation of the background in terms of Gaussian null coordinates,  which allow a non-singular description of the probe scalar field at the horizon. Upon Kaluza-Klein (KK) reduction on a respective sphere, the problem reduces to  examining the bulk massive wave equation on $AdS_{p+2}$ space by imposing boundary conditions. However, as we will see, by writing the background through Gaussian null coordinates, $AdS_{p+2}$ can be viewed as a conformal compactification of bi-dimensional anti de-Sitter space with a $p$-sphere, $AdS_2\times S^{p}$. Next, we shall discuss the bulk solutions through expanding the scalar field in modes, namely, normalizable and non-normalizable ones. From the AdS/CFT point of view, the non-normalizable modes are sources for the CFT operator living on the boundary of $AdS_{p+2}$ while normalizable modes vanish on the boundary. We will restrict ourselves to examine the normalizable modes which means that the boundary effects will drop out.  

This paper is organized as follows. In section \ref{sec:brane} we give an overview on extremal black $p$-branes from effective string theory. In section \ref{sec2}, we  focus on non-dilatonic extremal black $p$-branes and  we express their near-horizon geometries in Gaussian null coordinates, which provide a regular coordinate chart on the horizon. In subsection \ref{subsec1}, we probe the near-horizon geometry of the brane with a scalar field. Next, we expand the scalar field in modes fulfilling the isometries of the near horizon geometry. In subsection \ref{subsec2} we find explicit the Aretakis quantities on the horizon. In subsection \ref{subsec3}, we examine the  late time behavior of the scalar field in the $AdS_2$ section. In Appendix \ref{appendix1} we solve the bulk massless Klein-Gordon equation in Gaussian null coordinates taking into account the isometries of the near-horizon geometry. We find all regular solutions satisfying the boundary conditions, namely, normalizable and non-normalizable modes.  Finally, In Appendix \ref{appendix2} we display the quantization conditions to get regular solutions on the boundary of $AdS_{p+2}$.

\section{Extremal black p-brane ansatz}
\label{sec:brane}

Our starting point will be the relevant bosonic part of the $D$-dimensional effective Lagrangian arising from string theory \cite{Stelle:1996tz}: 

\begin{equation}
\mathcal{L}=e\left(R-\frac{1}{2}(\partial\phi)^2-\frac{1}{2(d+1)!}e^{a\phi}F_{d+1}^{2}\right),
\label{L1}
\end{equation} 
where $\phi$ is the dilaton, $F_{d+1}$ is the $d+1$-form field strength  corresponding to the $(d)$-form gauge field $A_{d}$, thus $F_{d+1}=dA_{d}$. The parameter $a$ controls the interaction between the dilaton and the field strength. 

The equations of motion  obtained varying 
Eq. (\ref{L1}) with respect to the dynamical fields  are
\begin{equation}
\begin{split}
& R_{MN}=\frac{1}{2}\partial_{M}\phi\partial_{N}\phi+\frac{1}{2(d)!}e^{a\phi}\big(F_{M...}F^{...} 	_{N}-\frac{d}{(d+1)(D-2)}g_{MN}F^2\big),\\
& \nabla_{K}(e^{a\phi}F^{K M_{1}...M_{N}})=0,\\
&\square\phi=\frac{a}{2(d+1)!}e^{a\phi}F^2.
\end{split}
\end{equation}

The equations of motion can be easily solved  after requiring symmetries of the metric with or
without  preserving some supersymmetries. In our case we require the metric ansatz to  possess a  $(\mbox{Poincar\'{e}})_{p+1}\times\mbox{SO}(D-(p+1)$) \cite{Duff:1996hp, Lu:1995yn}  symmetry. This leads to the spacetime  which can be interpreted as a $(p+1)$-dimensional hyperplane embedded in the whole $D$-dimensional manifold. A  suitable chart covering  the whole spacetime may be chosen  through splitting the coordinates into two components, namely, $x^{M}=(x^{\mu},y^{m})$, where the Greek letters label  coordinates which are ``parallel''  to the worldvolume, $(\mu=0,...,p=d-1)$, while the lowercase Latin letters label  coordinates which are ``transverse''  to the worldvolume  $(m=p+1,...,D-1)$.  In 
this notation, the ansatz for the metric takes the form
\begin{equation}
ds^2= e^{A(r)}\eta_{\mu\nu}dx^{\mu}dx^{\nu}+e^{B(r)}dy^{m}dy^{m},
\label{A1}
\end{equation}  
where $\eta_{\mu\nu}$ is the $(p+1)$-dimensional Minkowski metric, and $r=\sqrt{y^{m}y^{m}}$ is the radial coordinate in the transverse space. The former ansatz only deals with extremal black $p$-branes, which will be sufficient for our purposes in this work.  The ansatz is $(\mbox{Poincar\'{e}})_{p+1}\times\mbox{SO}(D-(p+1)$) invariant because the metric functions depend only on $r$. The corresponding ansatz for the dilaton and the gauge field requiring $(\mbox{Poincar\'{e}})_{p+1}\times\mbox{SO}(D-(p+1)$) symmetry  takes the form 
\begin{equation}
\begin{split}
&\phi=\phi(r),\\
&F_{m M_{1}...M_{p+1}}=\epsilon_{M_{1}...M_{p+1}}\partial_{m}e^{C(r)}\,\,\, \mbox{or} \,\,\,F_{N_{1}...N_{p+2}}=\epsilon_{N_{1}...N_{p+2}m}\frac{y^{m}}{r^{p+3}},
\label{A2}
\end{split}
\end{equation} 
	where $\epsilon_{M_{1}...M_{p+1}}$ is the volume $p$-form and the first and second field strengths describe the elementary $p$-brane and solitonic $p$-brane  (dual to the elementary case), respectively. 
	
	Substituting the ansatz in the field equations, one finds 
	\begin{eqnarray}
	\label{B1} &ds^2=H^{\frac{-4\tilde{d}}{\Delta(D-2)}}\eta_{\mu\nu}dx^{\mu}dx^{\nu}+H^{\frac{4d}{\Delta(D-2)}}dy^{i}dy^{i},\\
	&e^{\phi}=H^{\frac{2a}{\xi\Delta}},\\
	&H(r)=1+\frac{k}{r^{\tilde{d}}},
	\end{eqnarray}
where $\tilde{d}=D-d-2$ and $\xi=+1$ or $-1$ for the elementary and solitonic cases, respectively. For the explicit definitions of the function $C(r)$ and the parameter $k$, see \cite{Lu:1995yn}. The parameter $\Delta$ is related to the prefactor $a$  through the relation
\begin{equation}
a^2=\Delta-\frac{2d\tilde{d}}{D-2}.
\end{equation}
In order to get supersymmetric black $p$-branes one requires that the number $\Delta$ is related to the number of preserving supersymmetries $N$ by $N=\dfrac{4}{\Delta}$. Physically speaking, $N$ is an integer number corresponding to the number of strength fields were put in a compact form displayed in Eq.(\ref{L1}) once a particular basis has been chosen \cite{Klebanov:1996un}. In spite of this, black $p$-branes can exist for arbitrary $N$, even for non-integer values of $N$ which correspond to non-supersymmetric black $p$-branes. In next section we shall focus on a particular kind of black $p$-brane; non-dilatonic ($a=0$).

\section{The near-horizon limit}
\label{sec2}

 The aim of this section is 
 to find a near-horizon description for extremal non-dilatonic black branes, not necessarily supersymmetric 
  since we are considering arbitrary $N$.
  We concentrate our discussion on a particular sort of extremal non-dilatonic black $p$-branes discussed in \cite{Gibbons:1994vm} which are obtained by setting the dilaton to be zero, $a=0$, and $D=p+4$. Such a class of extremal black $p$-branes are interesting because they reduce upon dimensional reduction to extremal dilatonic black holes in four dimensions \cite{Gibbons:1994vm}. In the $p$-branes of our interest the metric and the field strength take respectively the form:  

\begin{equation}
ds^2=\bigg(1+\frac{k}{r}\bigg)^{-\frac{2}{p+1}}\bigg[(-dt^2+dx^a dx^a)+\bigg(1+\frac{k}{r}\bigg)^{\frac{2(p+2)}{p+1}}(dr^2+r^2 d\Omega^{2}_{2})\bigg],
\label{C1}
\end{equation}
\begin{equation}
F_{rM_{1}...M_{p+1}}=-\frac{2}{\sqrt{2}}\sqrt{\frac{p+2}{p+1}}\epsilon_{M_{1}...M_{p+1}}\frac{k}{r^2},
\end{equation}
where $a=1,...,p$. Note that the degenerate Killing horizon is located at $r=0$. Furthermore, the metric can be  rewritten in  Schwarzschild-like form by identifying $r=\tilde{r}-k$:
\begin{equation}
ds^2=\bigg(1-\frac{k}{\tilde{r}}\bigg)^{\frac{2}{p+1}}(-dt^2+dx^a dx^a)+\bigg(1-\frac{k}{\tilde{r}}\bigg)^{-2}d\tilde{r}^2+\tilde{r}^2 d\Omega^2_{2}.
\label{metric}
\end{equation} 
The horizon  is now located at $\tilde{r}=k$. In particular,  the case $p=0$ recovers  the extremal Reissner-Nordstr\"{o}m solution in GR. Henceforth, we can use the metric in the form (\ref{C1}). 

As aforementioned the extremal non-dilatonic black $p$-brane considered above can be viewed as an extremal dilatonic black hole in four dimensions. Upon carrying out the double-dimensional reduction from $D=p+4$ to $D=4$,  one finds a Lagrangian similar to Eq.(\ref{L1}) in four dimensions whose solution describes an extremal magnetically-charged black hole \cite{Garfinkle:1990qj} given by
\begin{equation}
\begin{split}
ds^{2}_{4}&=-\left(1+\frac{k}{r}\right)^{-\frac{p+2}{p+1}}dt^2+\left(1+\frac{k}{r}\right)^{\frac{p+2}{p+1}}\left(dr^2+r^2 d\Omega^{2}_{2}\right);\\
e^{a\phi}&=\left(1+\frac{k}{r}\right)^{\frac{p}{2(p+1)}};\\
F_2&=Q\epsilon_2,
\label{bh}
\end{split}
\end{equation}  
where the dilaton coupling takes a particular value $a^2=\dfrac{p}{(p+2)}$ and $\epsilon_2$ is the volume two-form. This setup mimic
s a non-dilatonic black $p$-brane as embedding in $(p+4)$-dimensional space-time. Moreover, the solution (\ref{bh}) has a BPS bound which relates the ADM mass $(M)$ and the charge $(Q)$ of the black hole to the constant $k$ by the equation
\begin{equation}
\frac{p+2}{2(p+1)}k=M\geq \sqrt{\frac{p+2}{2(p+1)}}Q.
\end{equation}
Note that in the non-dilatonic black $p$-brane description, the ADM mass and charge should be taken into account per unit volume of $p$-branes instead. For our purposes we have found more convenient to work with the extremal non-dilatonic black $p$-brane description since it has a simple near-horizon geometry, which is not the case for the extremal dilatonic black hole \cite{Gibbons:1994vm, Gibbons:1993sv}. On the face of it, both metrics are related to each other by
\begin{equation}
ds^{2}_{p+4}=\left(1+\frac{k}{r}\right)^{-\frac{2}{p+1}} dx^a dx^a + \left(1+\frac{k}{r}\right)^{\frac{p}{p+1}}ds^{2}_{4},
\label{p4}
\end{equation}  
 where $ds^{2}_{p+4}$ is the extremal non-dilatonic black $p$-brane space-time (\ref{C1}).

In order to examine the behavior of the metric (\ref{C1}) near the horizon, one needs a non-singular near-horizon description of it. In the limit $r\rightarrow 0$  it is easily seen that the near-horizon geometry of (\ref{C1}) is given by $AdS_{p+2} \times S^2$. More explicitly, let us  perform the following coordinate transformation: $\bigg(\dfrac{k}{r}\bigg)^{-\frac{1}{p+1}}=\lambda$. As a result, $\lambda\rightarrow 0$ near the horizon. Thus, the near-horizon metric  takes the form
\begin{equation}
ds^2_{NH}=\lambda^2\big[(-dt^2+dx^a dx^a)\big]+\frac{k^2(p+1)^2}{\lambda^2}d\lambda^2+k^2 d\Omega^2_{2},
\end{equation} 
which can be  rewritten in a more familiar form after  one more coordinate transformation $\lambda^{\prime}=(p+1)k\lambda$ and  defining $k^{\prime}=k(p+1)$, so the metric takes the form
\begin{equation}
ds^2_{NH}=\bigg(\frac{\lambda^{\prime}}{k^{\prime}}\bigg)^2\bigg[(-dt^2+dx^a dx^a)\bigg]+\bigg(\frac{k^{\prime}}{\lambda^{\prime}}\bigg)^2 d\lambda^{\prime 2}+k^2 d\Omega^2_{2}.
\end{equation}
Additionally,  by taking the following coordinate transformation $\tau=\dfrac{k^{\prime 2}}{\lambda^{\prime}}$ with $\tau>0$, we get
\begin{equation}
ds^2_{NH}=\frac{k^{\prime 2}}{\tau^2}\left[-dt^2+dx^a dx^a + d\tau^2\right]+k^2 d\Omega^2_{2},
\label{UHP}
\end{equation}
which is a product of $AdS_{p+2}$ with $S^2$. In the former equation, $AdS_{p+2}$ is represented in terms of the upper half plane representation. The $AdS$ radius is $k^{\prime}$ which in turn depends only on $k$ and the dimension of the space-time. The metric (\ref{C1})  is characterized by two distinct asymptotic behaviors  for large $r$ and near the horizon: it interpolates between the $D$-dimensional flat space and $AdS_{p+2}\times S^2$, when $r$ goes to infinity and in the near-horizon limit, respectively. A further  problem comes up since the metric in any of the coordinate systems introduced above is singular at the horizon $\tau=+\infty$ ($\lambda^{\prime}=0$) and therefore none of these coordinates are  appropriate to describe the near-horizon geometry.

To get a complete non-singular metric on the horizon, we should describe the near-horizon metric in Gaussian null coordinate
s \cite{Kunduri:2007vf}, assuming  that the new coordinates 
in the neighborhood on the horizon are $(v,r,y^{b},\Omega_{2})$, with ($b=1,...,p$)  parameterizes a particular $p$-dimensional spatial section of the full space. They are related to the Poincar\'{e} coordinates, $(t,\lambda^{\prime},x^{a},\Omega_{2})$, explicitly:
\begin{equation}
\lambda^{\prime}=r\cosh(\eta), \,\,\, t=\bigg(v+\frac{1}{r}\bigg)k^{\prime 2}, \,\,\, x^{a}=\frac{\mu^{a}\tanh(\eta)}{r}k^{\prime 2},
\end{equation}
where $0\leq\eta=y^{(1)}< +\infty$, $v$ is 
an ingoing (retarded) null coordinate and $\mu^{a}$ parametrize a sphere, $S^{p}$, 
 $\mu^{a}\mu^{a}=1$. In terms of the new coordinates the metric takes the form
\begin{equation}
ds^{2}_{NH}=k^{\prime 2}\bigg[\cosh^2(\eta)\bigg(-r^{2}d v^2 + 2dv dr\bigg)+d\eta^2 +\sinh^2(\eta) d\Omega^{2}_{p-1}\bigg]+k^2 d\Omega^{2}_{2},
\label{M1}
\end{equation} 
 which obviously  describes the space $AdS_{p+2} \times S^2$. In fact, the metric in Gaussian null coordinate is completely non-singular 
 at the horizon $r=0$, and is extended to the region behind the horizon
 through the reflection invariance of the full metric
\begin{equation}
\begin{split}
r&\rightarrow-r;\\
v&\rightarrow-v.
\end{split}
\end{equation}  

\begin{figure}
					\centering
							\includegraphics[width=0.65\textwidth]{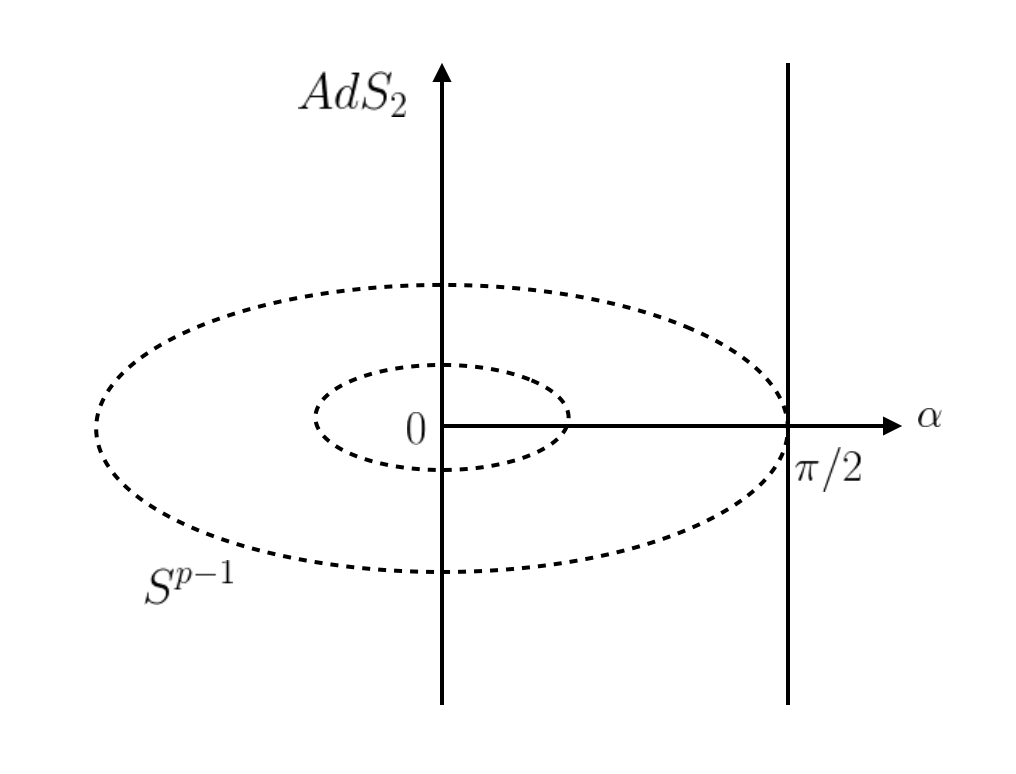}
					\caption{\label{topology} The figure displays the topology of $AdS_{p+2}$. The circles stand for $S^{p-1}$-spheres whose radii are $\alpha$ and the boundary is located at $\alpha=\pi/2$ axis. The vertical axis consists of copies of $AdS_2$.}
				\end{figure}

 We now observe that the former metric can be seen as a warped product of $AdS_{2}$ with a $p$-dimensional hyperbolic space $\mathcal{H}^p$  plus  $S^2$. It is worth pointing out that Eq. (\ref{M1}) is singular at $\eta=+\infty$.
However, let us take the 
coordinate transformation $\tan(\alpha)=\sinh(\eta)$ in Eq. (\ref{M1}) with $0\leq \theta<\pi/2$. 
We  get
\begin{equation}
ds^{2}_{NH}=\frac{k^{\prime 2}}{\cos^2(\alpha)}\bigg[\bigg(-r^{2}d v^2 + 2dv dr\bigg)+d\alpha^2 +\sin^2(\alpha) d\Omega^{2}_{p-1}\bigg]+k^2 d\Omega^{2}_{2},
\label{cos}
\end{equation}
where the geometry is clearly $AdS_{p+2}\times S^2$ as expected. However, now the first term in the former metric is can be viewed as a conformal compactification of a lower-dimensional anti-de Sitter space, $AdS_{2}$, times a $p$-sphere, $S^{p}$. The conformal boundary of $AdS_{p+2}$ is defined by the divergence of the conformal factor $\frac{k^{\prime 2}}{\cos^2(\alpha)}$ at  $\alpha=\frac{\pi}{2}$. In Gaussian null coordinates the conformal boundary takes the  form $AdS_2 \times S^{p-1}$, which is isometric to the $p+1$-dimensional Minkowski space. The $AdS_{p+2}$ topology in Gaussian null coordinates  is schematically presented in Fig.(\ref{topology}).

It is interesting to note that the near-horizon geometry of the extremal non-dilatonic black $p$-brane written in the form given by Eq.(\ref{cos}) shows ``external'' and ``internal'' structures. Looking at Fig.(\ref{topology}), the external structure displays the behavior of $AdS_{p+2}$ in terms of $p$-extra spatial dimensions which has its own boundary attained at $\alpha=\pi/2$. It turns out that the vertical axis in Fig.(\ref{topology}) represents the $AdS_2$, a section of the thorough space. Thus, the internal structure of the $AdS_{2}$ section, which has its own horizon located at $r=0$, is warped by the $\alpha$-coordinate as we have said previously. We are dealing in this work with near-horizon geometries, so it is enough to restrict our analysis to $r\rightarrow 0$ . From a physical perspective, $AdS_{p+2}$ is nothing more than a warped geometry of $AdS_{2}\times S^p$: the $AdS_{p+2}$ radius is no longer a constant, but depends on the radial coordinate of the compact space $S^{p-1}$ instead. This is not a novel property since many examples of warped geometries arises from dimensional reduction in string theory and supergravity \cite{Maldacena:1998uz, vanNieuwenhuizen:1984ri}. In particular, the warped $AdS_{3}$ is built up as a Hopf fibration over $AdS_{2}$ \cite{ Anninos:2008qb,Castro:2014ima}. Hence, the near-horizon geometry of extremal dilatonic black holes Eq.(\ref{bh}) 
is reached upon the double-dimensional reduction of extremal non-dilatonic black $p$-branes near the horizon by using Eq.(\ref{p4}),
and their forms are conformal deformations of $AdS_{2}\times S^2$. More explicitly, by doing this we 
find
\begin{equation}
ds^2_{NH}=\Omega(r)\left[k^{\prime 2}\left(-r^2 dv^2+ 2dv dr\right) +k^2 d\Omega_{2}^{2}\right],
\end{equation}
where the conformal factor is $\Omega(r)=\left(\dfrac{r}{k^{\prime}}\right)^p$ and the metric inside the brackets is $AdS_2\times S^2$. 
 For example, if we take $p=0$ we recover the near-horizon geometry of the extremal RN black hole which is just $AdS_2 \times S^2$, though for $p\neq 0$ the remaining geometry is a deformation of $AdS_2\times S^2$ as we have seen by re-interpreting the black hole as a black $p$-brane in higher dimensions.

It is convenient to  rewrite the metric (\ref{cos}) in a more compact fashion:
\begin{equation}
\begin{split}
ds^{2}_{NH}&=\frac{k^{\prime 2}}{\cos^2(\alpha)}\bigg[\bigg(-r^{2}d v^2 + 2dv dr\bigg)+d\Omega^{2}_{p}\bigg]+\gamma_{IJ}dx^I dx^J\\
&=\frac{k^{\prime 2}}{\cos^2(\alpha)}\bigg[\bigg(-r^{2}d v^2 + 2dv dr\bigg)+\omega_{ab}dy^{a}dy^{b}\bigg]+\gamma_{IJ}dx^I dx^J,
\label{NHm}
\end{split}
\end{equation}
where $d\Omega^{2}_{p}=d\alpha^2 +\sin^2(\alpha) d\Omega^{2}_{p-1}$ and, $\omega_{ab}$ and $\gamma_{IJ}$ represent the $p$-dimensional  spherical metric and $2$-dimensional metric, respectively. The horizon is a degenerate Killing horizon which is a null hypersurface defined by the vanishing of the norm of the Killing vector $\dfrac{\partial}{\partial v}$. At the horizon the induced metric looks like
\begin{equation}
ds^{2}_{H}=\frac{k^{\prime 2}}{\cos^2(\alpha)}\omega_{ab}dy^{a}dy^{b}+\gamma_{IJ}dx^I dx^J,
\label{horizon}
\end{equation}
where the first term represents a $p$-dimensional hyperbolic manifold $\mathcal{H}^p$ and the second one stands for a $S^2$. So, black $p$-brane horizons are non-compact surfaces. 

 $AdS$ spaces with a conformal boundary are not globally hyperbolic \cite{Ishibashi:2003jd,Ishibashi:2004wx}. Therefore in order to define a  well-posed Cauchy problem, besides  of picking adequate initial data, it is necessary  to impose suitable boundary conditions at the conformal boundary to guarantee smooth solutions of the Klein-Gordon and/or Dirac equations in the bulk. In the next section we shall examine a scalar  field on this background prescribing further boundary conditions.

\subsection{Scalar field}
\label{subsec1}

In this section, we consider a probe massless scalar field minimally coupled to  gravity for the metric given by Eq. (\ref{NHm}). The field action is given by
	\begin{equation}
S[\psi]=\frac{1}{2}\int d^{D}x\sqrt{-g}\left\{\partial_{A}\psi\partial^{A}\psi\right\}.
\label{action}
\end{equation}	
Since the $AdS$ space is not compact ($AdS$ has an infinite volume), the action in principle diverges. So, in order to find finite physical quantities one needs a regularization of the volume to avoid divergent terms \cite{Polchinski:2010hw}, as we shall see below. Varying the scalar field action with respect to $\psi$, we have
\begin{eqnarray}
\label{kj}
	0&=&\int d^{p+1}y\int_{0}^{\pi/2 -\epsilon} d\alpha \int d^2 x \sqrt{-g}\bigg[ g^{AB} (\partial_{A}\psi)\partial_{B}\psi\bigg] \\
	&=&\int d^2 x\int d^{p+1}y\int_{0}^{\pi/2 -\epsilon} d\alpha\left\{\partial_{A}\bigg(\sqrt{-g} g^{AB}\partial_{B}\psi\delta\psi\bigg)-\bigg[\partial_{A}\bigg(\sqrt{-g}g^{AB}\partial_{B}\psi\bigg)\bigg]\delta\psi\right\},\nonumber
\end{eqnarray}	
  where we have used $\epsilon$ as a cutoff parameter to regularize the infinite volume of $AdS_{p+2}$ and  integrated by parts from the first to second line.  Apart from this, the bulk has been restricted temporarily to the region $0\leq \alpha\leq\ \pi/2 -\epsilon$. Expanding all terms in Eq. (\ref{kj}), we  arrive at
	\begin{eqnarray}
	\nonumber 0&=&\int d^2 x\int_{AdS_{p+2}} d^{p+2}y \bigg[-\left\{\frac{1}{\sqrt{-g}}\partial_{A}\bigg(\sqrt{-g}g^{AB}\partial_{B}\psi\bigg)\right\}\sqrt{-g}\delta\psi\bigg]+\\
	&+&\int d^2 x\int d^{p+1}y \sqrt{-h}\,n^{\alpha}\partial_{\alpha}\psi\,\delta\psi\Big|_{\alpha=0}^{\alpha=\pi/2-\epsilon},
	\label{KGE}
	\end{eqnarray}
 where  $h$ is the determinant of the induced metric	$h_{\mu\nu}$ defined at the boundary, $n^{\alpha}$ is a unit normal vector to the boundary
 , and the first integral is performed in the bulk whilst the second integral is performed  between the initial and end points of $\alpha$. The first term in Eq.(\ref{KGE}) is  proportional to the Klein-Gordon equation,  and the second one is a boundary term. As usual, in order to get a well-defined variation principle,  the boundary term must vanish for variations  obeying special boundary conditions.  Afterwards, the problem reduces to  solving the Klein-Gordon equation in the bulk for particular boundary conditions  which enforce the boundary term  to vanish.
	
	Now, let us solve the Klein-Gordon equation on the background given by Eq. (\ref{NHm}). To  start, we write 
\begin{equation}
\square \psi=\frac{1}{\sqrt{-g}}\partial_{M}\left(\sqrt{-g}g^{MN}\partial_{N}\psi\right)=0,
\label{KG}
\end{equation} 
where $\sqrt{-g}=\bigg(\dfrac{k^{\prime}}{\cos\alpha}\bigg)^{p+2}\sqrt{\omega}\sqrt{\gamma}$. The explicit form of the metric is as follows:
\begin{eqnarray}
(g_{MN})=\left(\begin{array}{cccc}
-\dfrac{r^2 k^{\prime 2}}{\cos^2(\alpha)} & \dfrac{k^{\prime 2}}{\cos^2(\alpha)} & 0 & 0\\
\dfrac{k^{\prime 2}}{\cos^2(\alpha)} & 0 & 0 & 0\\
0 & 0 & \dfrac{k^{\prime 2}}{\cos^2(\alpha)}\omega_{ab} & 0\\
0 & 0 & 0 & \gamma_{IJ},
\end{array}\right),
\end{eqnarray}

\begin{eqnarray}
(g^{MN})=\left(\begin{array}{cccc}
0 & \dfrac{\cos^2(\alpha)}{k^{\prime 2}} & 0 & 0\\
\dfrac{\cos^2(\alpha)}{k^{\prime 2}} & \dfrac{r^2 \cos^2(\alpha)}{k^{\prime 2}} & 0 & 0\\
0 & 0 & \dfrac{\cos^2(\alpha)}{k^{\prime 2}}\omega^{ab} & 0\\
0 & 0 & 0 & \gamma^{IJ},
\end{array}\right),
\label{matrixg}
\end{eqnarray}
where $\omega^{ab}$ and $\gamma^{IJ}$ represent the inverse of $\omega_{ab}$ and $\gamma_{IJ}$, respectively. As  we  noted earlier, our aim is to find possible  conserved charges in the near-horizon limit. Therefore, 
Eq. (\ref{KG}) reduces to
\begin{equation}
\begin{split}
0&=\frac{1}{\cos^{(p)}(\alpha)}\sqrt{\omega}\sqrt{\gamma}2\partial_{v}\left[\partial_{r}\psi\right]+\frac{1}{\cos^{(p)}(\alpha)}\sqrt{\omega}\sqrt{\gamma}\partial_{r}\left[r^2\partial_{r}\psi\right]\\
&+\sqrt{\gamma}\partial_{a}\left[\frac{1}{\cos^{(p)}(\alpha)}\sqrt{\omega}\omega^{ab}\,\partial_{b}\psi\right]+\frac{k^{\prime 2}}{\cos^{(p+2)}(\alpha)}\sqrt{\omega}\,\partial_{I}\left[\sqrt{\gamma}\gamma^{IJ}\partial_{J}\psi\right],
\label{kgop}
\end{split}
\end{equation}
or, more explicitly,
\begin{eqnarray}
 0&=&\nonumber \frac{\sin^{(p-1)}(\alpha)}{\cos^{(p)}(\alpha)}2\sqrt{\omega_{(p-1)}}\sqrt{\gamma}\partial_{v}\left[\partial_{r}\psi\right]+ \frac{\sin^{(p-1)}(\alpha)}{\cos^{(p)}(\alpha)}\sqrt{\omega_{(p-1)}}\sqrt{\gamma}\partial_{r}\left[r^2\partial_{r}\psi\right]\\
\nonumber &+&\sqrt{\omega_{(p-1)}}\sqrt{\gamma}\partial_{\alpha}\left[\frac{\sin^{(p-1)}(\alpha)}{\cos^{(p)}(\alpha)}\partial_{\alpha}\psi\right]+k^{\prime 2}\frac{\sin^{(p-1)}(\alpha)}{\cos^{(p+2)}(\alpha)}\sqrt{\omega_{(p-1)}}\,\partial_{I}\left[\sqrt{\gamma}\gamma^{IJ}\partial_{J}\psi\right]\\
&+&\frac{\sin^{(p-3)}(\alpha)}{\cos^{(p)}(\alpha)}\sqrt{\gamma}\,\partial_{i}\left[\sqrt{\omega_{(p-1)}}\omega^{ij}_{(p-1)}\partial_{j}\psi\right],
\label{KGOp}
\end{eqnarray}
where $\omega_{(p-1)}$ is the determinant of the $(p-1)$-dimensional  spherical metric $\omega^{ij}_{(p-1)}$, and $i,j=1...p$. We propose as an ansatz the following expansion in modes: 
\begin{equation}
\psi=\sum_{l,l^{\prime},m} \Phi_{m}(v,r)Y_{l^{\prime}}(\Omega_{(p-1)})Y_{l}(\Omega_{2})\phi_{l^{\prime}m}(\alpha),
\label{AN}
\end{equation}
where $Y_{l}(\Omega_{2})$ and $Y_{l^{\prime}}(\Omega_{(p-1)})\equiv  Y_{l_{1},...,l_{p-1}}(\Omega_{(p-1)})$ are the spherical harmonics on $S^2$ and $S^{(p-1)}$
, their quantum numbers  being  respectively $l$, $l^{\prime}$.  More generically, $l_{1},...,l_{p-1}$ are  integer numbers satisfying the following condition:
\begin{equation}
l_{p-1}\geq l_{p-2}\geq ...\geq l_{2} \geq |l_{1}|.
\end{equation} 
The spherical harmonics  obey the equations \cite{Ammon:2015wua}:
\begin{eqnarray}
\label{SH1}\square_{S^2} Y_{l}(\Omega_{2})&=&-l(l+1)Y_{l}(\Omega_{2}),\\
\label{SH2}\square_{S^{(p-1)}} Y_{l^{\prime}}(\Omega_{(p-1)})&=&-l^{\prime}(l^{\prime}+p-2)Y_{l^{\prime}}(\Omega_{(p-1)}),
\end{eqnarray}
where we have used the shorthand notation $\sum_{l^{\prime}}\equiv\sum_{l_{p-1}=1}^{\infty}\sum_{l_{p-2}=1}^{l_{p-1}}...\sum_{l_{1}=-l_{2}}^{l_{2}}$. The quantum number $m$ stems from the fulfillment of regularity conditions of the radial solutions ($\phi(\alpha)$) at the origin and boundary (see Appendix \ref{appendix1}); in other words, $m$ is the radial quantum number.

Substituting Eqs. (\ref{SH1}-\ref{SH2}-\ref{AN}) into Eq. (\ref{KGOp}), we have
\begin{eqnarray}
\nonumber 0&=& \frac{\sin^{(p-1)}(\alpha)}{\cos^{(p)}(\alpha)}\frac{\sqrt{\omega_{(p-1)}}\sqrt{\gamma}}{\Phi(v,r)}\left[2\partial_{v}\partial_{r}\Phi(v,r)+\partial_{r}\left(r^2\partial_{r}\Phi(v,r)\right)\right]+\\
\nonumber&+&\sqrt{\omega_{(p-1)}}\sqrt{\gamma}\frac{1}{\phi(\alpha)}\partial_{\alpha}\left[\frac{\sin^{(p-1)}(\alpha)}{\cos^{(p)}(\alpha)}\partial_{\alpha}\phi(\alpha)\right]-m^{2}_{0}\frac{\sin^{(p-1)}(\alpha)}{\cos^{(p+2)}(\alpha)}\sqrt{\omega_{(p-1)}}\sqrt{\gamma}+\\
&+&\frac{\sin^{(p-3)}(\alpha)}{\cos^{(p)}(\alpha)}\sqrt{\omega_{(p-1)}}\sqrt{\gamma}\,l^{\prime}(l^{\prime}+p-2),
\label{KGOP1}
\end{eqnarray}
where we defined a ``mass" term\footnote{Note that this term could be negative.}, $m^{2}_{0}=(p+1)^2 l(l+1)$, and omitted the index of the functions for convenience. Notice that this term comes up after Kaluza-Klein (KK) reduction on $S^2$. The former equation can be split into two decoupled differential equations: the first one, a ordinary differential equation for the radial function and the second, a partial differential equation for the $AdS_2$-section function:
\begin{eqnarray}
\nonumber 0&=&\partial_{\alpha}\left[\frac{\sin^{(p-1)}(\alpha)}{\cos^{(p)}(\alpha)}\partial_{\alpha}\phi(\alpha)\right]+\bigg[\xi^2 \frac{\sin^{(p-1)}(\alpha)}{\cos^{(p)}(\alpha)}-m^{2}_{0}\frac{\sin^{(p-1)}(\alpha)}{\cos^{(p+2)}(\alpha)}-\\
\label{KG1}&-&l^{\prime}(l^{\prime}+p-2)\frac{\sin^{(p-3)}(\alpha)}{\cos^{p}(\alpha)}\bigg] \phi(\alpha),\\
0&=&2\partial_{v}\left[\partial_{r}\Phi(v,r)\right]+\partial_{r}\left[r^2\partial_{r}\Phi(v,r)\right]-\xi^2 \Phi(v,r),
\label{vr}
\end{eqnarray}
with $\xi^2$ a separation constant. The explicit solutions for the radial equation are displayed in Appendix \ref{appendix1}. Here it should be stressed that $\xi^2$ (see Appendix \ref{appendix2}) is a combination of the set of quantum numbers $\left\{l, l^{\prime}, m\right\}$. As discussed in \cite{Balasubramanian:1998sn} for  $AdS$ spaces, the general solution for the scalar field is suitably split into two parts: the normalizable  and non-normalizable modes, such modes are explicitly obtained in Appendix \ref{appendix1} for $AdS$ in  Gaussian null  coordinates. The $AdS_{2}$-section equation is discussed in more details in Section \ref{subsec3}.

\subsection{Conserved charge on the horizon} 
\label{subsec2}

In this subsection we shall discuss the existence of conserved charges on the horizon of  black branes. Similarly to the context of black holes in GR where conserved charges on the horizon have been already obtained in \cite{Bizon:2012we,Godazgar:2017igz}, it is reasonable to expect such conserved charges  to hold for black branes as well. We start by considering the wave equation \ref{KGOp} along with the ansatz Eq.(\ref{AN}). Upon integration over the compact manifolds $S^2$ and $S^{p-1}$, the last two terms in it vanish as a consequence of spherical harmonics properties. Now, integrating the remaining equation over the ``extra'' dimension $\alpha$ and evaluating at the horizon, we obtain 
\begin{equation}
			0=\int d V_{\mathcal{H}^{p}}\int d^{2}x\,\sqrt{\gamma}\bigg[2\partial_{v}\partial_{r}\psi+\mathcal{A}\psi\bigg],
			\label{horizon1}
			\end{equation}
where $V_{\mathcal{H}^{p}}$ stands for the volume of the $p$-dimensional hyperboloid and $\mathcal{A}$ is an operator defined by:
\begin{equation}
\mathcal{A}=\frac{\cos^{(p)}(\alpha)}{\sin^{(p-1)}(\alpha)}\partial_{\alpha}\bigg(\frac{\sin^{(p-1)}(\alpha)}{\cos^{(p)}(\alpha)}\partial_{\alpha}\bigg).
\end{equation}
 Note that the Eq.(\ref{horizon1}) is integrated over the horizon as one can see looking at the Eq.(\ref{horizon}). Recalling that the horizon is a hyperbolic space plus a sphere, 
 we shall proceed with a regularization scheme since the horizon is non-compact. The regularized volume of the hyperboloid may be written in terms of the volume of the $(p-1)$-dimensional sphere \cite{Maldacena:2012xp}, namely, $V_{\mathcal{H}^{p}}^{\mbox{reg.}}=V_{S^{p-1}}\int_{0}^{\pi/2 -\epsilon}d\alpha\,\dfrac{\sin^{(p-1)}(\alpha)}{\cos^{(p)}(\alpha)}$, with $\epsilon$ is a cutoff  introduced in order to get finite results. Introducing   
  for convenience the shorthand notation $\psi=\sum\Psi(X)\phi(\alpha)$ (we are omitting the indices in the sum), with $X$ 
  standing for the other coordinates, it is possible 
  to carry out the integration over $\alpha$ by using the asymptotic behavior of bulk modes. An in-depth assessment of bulk modes of the wave equation in Gaussian null coordinates is provided in 
   Appendices \ref{appendix1}-\ref{appendix2}. Firstly, we shall explicit the bulk modes for a particular case where the parameter $\nu$ is a non-integer number\footnote{
   See Appendix \ref{appendix1} for the definition of $\nu$.}.  
   In this case they are expanded in terms of hypergeometric functions
\begin{eqnarray}
\nonumber\phi(\alpha)&=&C \sin^{l^{\prime}}(\alpha)\cos^{2\lambda_{+}}(\alpha)\,{_2}F_{1}\left(e_{+},f_{+};(1+\nu);\cos^{2}(\alpha)\right)+\\
&+&D \sin^{l^{\prime}}(\alpha)\cos^{2\lambda_{-}}(\alpha)\,{_2}F_{1}\left(e_{-},f_{-};(1-\nu);\cos^{2}(\alpha)\right),
\label{fc1}
\end{eqnarray}
where all the constants are defined in Appendix \ref{appendix1}. As well-known the hypergeometric functions form representations of the conformal group $SL(2,\mathbb{R})$. As a consequence, we can truncate the series to the leading terms near the boundary, they fall off as $\cos^{2\lambda_{\pm}}(\alpha)$. Therefore, the full solution satisfies the following asymptotic 
behavior,
\begin{equation}
\psi(X,\alpha)\sim\sum_{\left\{l,l^{\prime},m\right\}}\Psi(X)\cos^{2\lambda_{+}}(\alpha)+\gamma(X)\cos^{2\lambda_{-}}(\alpha),
\end{equation} 
near the boundary. In order to have well-behaved solutions at the boundary one needs 
to 
 impose boundary conditions (as displayed in Eqs.(\ref{QC1}, \ref{QC2})) which lead to $\Psi(X)$ and $\gamma(X)$ modes to be quantized. The standard quantization  condition 
corresponds to fixing $\gamma(X)=0$ or, 
equivalently, to taking $D=0$ as discussed in Appendix \ref{appendix1} and \ref{appendix2}. As a result, $\Psi(X)$ are normalizable quantized modes satisfying the quantization condition (\ref{QC1}). On the other hand, there is an alternative quantization condition which corresponds to fixing $\Psi(X)=0$ or, equivalently, to picking $C=0$ as discussed in Appendix \ref{appendix1} and \ref{appendix2}. In this case, $\gamma(X)$ are non-normalizable modes satisfying the quantization condition (\ref{QC2}). Of course, the above analysis only makes sense for $\nu>1$; otherwise, both modes are normalizable and a more careful analysis must be done to choose the fluctuation 
 as pointed out in Appendix \ref{appendix2}.

  The solution (\ref{fc1}) may be written in a more convenient way after applying the quantization conditions $(\ref{QC1}, \ref{QC2})$:
	\begin{eqnarray}
\nonumber\phi_{1}(\alpha)&=&C \frac{m!}{\left(2\lambda_{+}-\frac{(p-1)}{2}\right)_m}\sin^{l^{\prime}}(\alpha)\cos^{2\lambda_{+}}(\alpha) P_{m}^{(2\lambda_{+}-\frac{(p+1)}{2}, \frac{p}{2}+l^{\prime}-1)} (1-2\cos^{2}(\alpha))+\\
&+&D \frac{m!}{\left(2\lambda_{-}-\frac{(p-1)}{2}\right)_m}\sin^{l^{\prime}}(\alpha)\cos^{2\lambda_{-}}(\alpha) P_{m}^{(2\lambda_{-}-\frac{(p+1)}{2}, \frac{p}{2}+l^{\prime}-1)} (1-2\cos^{2}(\alpha))
\label{KC}
\end{eqnarray}
  where $P_{m}^{(a, b)}(x)$ are Jacobi polynomials, the quantity $(a)_{m}$ is defined in Eq.(\ref{pol}) and, $C$ and $D$ are constants given by Eqs.(\ref{EqCD},\ref{EqCD1}). In addition, $\lambda_{\pm}$ are the roots given by Eq.(\ref{lambda1}) defined from the hypergeometric equation (\ref{hyper}). Hence, in this case, the full solution may be conveniently written in a more appropriate form as follows:
	\begin{equation}
\phi_{1}(\alpha)=C^{(+)}\phi^{(+)}(\alpha)+C^{(-)}\phi^{(-)}(\alpha),
\end{equation}
where 
\begin{equation}
\begin{split}
C^{(+)}&=C \frac{m!}{\left(2\lambda_{+}-\frac{(p-1)}{2}\right)_m},\\
C^{(-)}&=D \frac{m!}{\left(2\lambda_{-}-\frac{(p-1)}{2}\right)_m},\\
\phi_{\pm}(\alpha)&=\sin^{l^{\prime}}(\alpha)\cos^{2\lambda_{\pm}}(\alpha) P_{m}^{(2\lambda_{\pm}-\frac{(p+1)}{2}, \frac{p}{2}+l^{\prime}-1)} (1-2\cos^{2}(\alpha)).
\end{split}
\end{equation}

The function $\phi^{(+)}(\alpha)$ is used to build the pure normalizable modes for all non-integer $\nu$ since the Jacobi polynomials are orthonormal the by applying the inner product \ref{norm}. The other function $\phi^{(-)}$ only can be cast into pure normalizable modes for the non-integer $\nu$ within the range $\nu<1$, in this way, the Jacobi polynomials can be made orthonormal within this range.

Now let us consider to the other case which corresponds to 
 taking $\nu$ to be an integer number. In this situation the solution is given by:
\begin{eqnarray}
\nonumber\phi_1 (\alpha)&=&D\sin(\alpha)^{l^{\prime}}\cos(\alpha)^{2\lambda_{-}}\sum_{k=0}^{\nu-1}\frac{\left(a-\nu\right)_{k} \left(b-\nu\right)_{k}}{k! \left(1-\nu\right)_k}\cos(\alpha)^{2k}-\\
\nonumber&-&C\sin(\alpha)^{l^{\prime}}\cos(\alpha)^{2\lambda_{+}}\sum_{k=0}^{\infty}\frac{\left(a\right)_k \left(b\right)_k}{k!(k+\nu)!}\cos(\alpha)^{2k}\bigg[\ln(\cos^{2}(\alpha))-\\
\nonumber&-&\psi(k+1)-\psi(k+\nu+1)+\psi\left(a+k\right)+\\
&+&\psi\left(b+k\right)\bigg],
\label{sc11}
\end{eqnarray}
where now $C$ and $D$ are constants given by Eqs.(\ref{eq11},\ref{eq12}).  As 
in the former case we select the pure non-normalizable modes by taking $C=0$, which is the quantization condition (\ref{QC2}). These modes falls off as $\cos^{2\lambda_{-}}(\alpha)$ near the boundary as expected to such modes. On the other hand, the normalizable modes are picked by taking $D=0$ which is the quantization condition (\ref{QC1}). However, one must be careful to do so since for these modes 
  poles of the gamma function are involved. 
Upon imposing the normalizable boundary conditions, we have:
\begin{eqnarray}
\nonumber\phi_1 (\alpha)&=&-\frac{(-1)^{\nu}\Gamma\left(\frac{p}{2}+l^{\prime}\right)}{\Gamma(-m-\nu)\Gamma(\frac{p}{2}+l^{\prime}+m)}\sin^{l^{\prime}}(\alpha)\cos^{2\lambda_{+}}(\alpha)\sum_{k=0}^{\infty}\frac{\left(-m\right)_k \left(\frac{p}{2}+l^{\prime}+m+\nu\right)_k}{k!(k+\nu)!}\times\\
\nonumber&\times& \cos^{2k}(\alpha)\bigg[\ln\left(\cos^2(\alpha)\right)-\psi(k+1)-\psi(k+\nu+1)+\psi(-m+k)+\psi(b+k)\bigg],\\
\label{sc}
\end{eqnarray}
where we have used Eq.(\ref{eq11}) 
for $C$. Note that the first gamma function in the denominator has negative integer numbers in its argument; at these points the reciprocal gamma function is zero, i.e., $\dfrac{1}{\Gamma(-m-\nu)}=0$. On the other hand, 
 the third $\psi$ function inside the sum has poles for negative integer number argument; so, this term is the only surviving one inside the sum while the others do not 
 contribute. The remaining term is:
	\begin{eqnarray}
	\nonumber\phi_1 (\alpha)&=&\frac{(-1)^{\nu}\Gamma\left(\frac{p}{2}+l^{\prime}\right)}{\Gamma(-m-\nu)\Gamma(\frac{p}{2}+l^{\prime}+m)}\sin^{l^{\prime}}(\alpha)\cos^{2\lambda_{+}}(\alpha)\sum_{k=0}^{\infty}\frac{\left(-m\right)_k \left(\frac{p}{2}+l^{\prime}+m+\nu\right)_k}{k!(k+\nu)!}\times\\
&\times& \cos^{2k}(\alpha)\bigg[\psi(-m+k)\bigg],
	\label{phii}
	\end{eqnarray}
	 We should extract  
	 a   finite limit from the term $\dfrac{\psi(-m+k)}{\Gamma(-m-\nu)}$
	 . Let us expand both functions around negative integer numbers using the regulator $\varepsilon$, namely:
	\begin{equation}
	\begin{split}
	\psi(-m+k+\varepsilon)&=-\frac{1}{\varepsilon}-\gamma+H_{(m-k)}+\sum_{j\geq 1}\bigg[H_{(m-k)}^{j+1}+(-1)^{k+1}\zeta(j+1)\bigg]\varepsilon^j;\\
	\Gamma(-m-\nu+\varepsilon)&=\frac{(-1)^{m+\nu}}{(m+\nu)!}\bigg[\frac{1}{\varepsilon}+\psi(m+\nu+1)+\frac{\varepsilon}{2}\bigg(\frac{\pi^2}{3}+\psi(m+\nu+1)^2-{\psi}^{\prime}(m+\nu)\bigg)+\mathcal(O)(\varepsilon^2)\bigg],
	\end{split}
	\end{equation}
	where $H_{m}$ are harmonic numbers and $\zeta(s)$ is the Riemann zeta function. Expanding the reciprocal gamma function and upon taking the limit $\varepsilon\rightarrow 0$ we have the regularized expression:
	\begin{equation}
	\lim_{\varepsilon \to 0}\frac{\psi(-m+k+\varepsilon)}{\Gamma(-m-\nu+\varepsilon)}=\frac{\psi(-m+k)}{\Gamma(-m-\nu)}=\frac{(m+\nu)\Gamma(m+\nu)}{(-1)^{m+\nu-1}}.
	\end{equation}
	Inserting this result into Eq.(\ref{phii}) one finds:
	\begin{eqnarray}
	\nonumber\phi_1 (\alpha)&=&\frac{(-1)^{1-m}(m+\nu)\Gamma\left(\frac{p}{2}+l^{\prime}\right)\Gamma\left(m+\nu\right)}{\Gamma(\frac{p}{2}+l^{\prime}+m)}\sin^{l^{\prime}}(\alpha)\cos^{2\lambda_{+}}(\alpha)\sum_{k=0}^{\infty}\frac{\left(-m\right)_k \left(\frac{p}{2}+l^{\prime}+m+\nu\right)_k}{k!(k+\nu)!}\times\\
&\times& \cos^{2k}(\alpha).
	\end{eqnarray}
Now it is easy to check that the asymptotic behaviour near the boundary for this case goes as $\cos^{2\lambda_{+}}(\alpha)$ as expected.

  For our purposes we shall select the pure normalizable modes for all $\nu$ which in turn are arrived at imposing the quantization condition Eq.(\ref{QC1}). Such a choice enforces the vanishing of the second term in Eq.(\ref{horizon1}) evaluated at the boundary. Physically speaking, invoking the AdS/CFT dictionary, only the non-normalizable or non-fluctuating modes survive at the boundary 
(and  are identified as sources for  dual CFT operators $\mathcal{O}$ living on the boundary) while the normalizable or 
 fluctuating modes vanish at the boundary. The only surviving term in Eq.(\ref{horizon1}) implies a conserved charge on the horizon, which we will call Aretakis charge due to its similarity to the black hole context:
\begin{equation}
H^{Aretakis}=\int d V_{\mathcal{H}^{p}}\int d^{2}x\,\sqrt{\gamma}\,2\partial_{r}\psi.
\label{AREe}
\end{equation}
Note that this quantity is finite since $\psi$ should be restricted to the normalizable modes. To proceed further
we will define mode-by-mode Aretakis quantities in terms of the quantum numbers. To do so we can split our analysis into two parts: the first one, for $p$ odd and $\nu$ either integer or null; the second one, for $p$ even and $\nu$ non-integer. The separation constant $\xi$ given by the Eq.(\ref{QC1}) for normalizable modes for all $\nu$ may be rewritten in a 
simpler way, namely, defining the new quantum number
 \begin{eqnarray}
L&=&2m+l^{\prime}+2\lambda_{+}-1\\
&=&2m+l^{\prime}+p+(p+1)l,\,\,\, \mbox{with}\,\,\, L=p,p+1,... ,
\end{eqnarray}
 where we have used the definition $m^{2}_{0}=(p+1)^{2}l(l+1)$ in the second line in the above equation. The quantum number $L$ is a combination of the other ones so that we find $\xi^2=L(L+1)$. Plugging it in Eq.(\ref{vr}) 
 we obtain:
\begin{equation}
0=2\partial_{v}\left[\partial_{r}\Phi(v,r)\right]+\partial_{r}\left[r^2\partial_{r}\Phi(v,r)\right]-L(L+1) \Phi(v,r),
\label{kgvr}
\end{equation}
where we have omitted the indices for convenience. 
Now by applying the operator $\partial^{L}_{r}$ in the above equation and using mathematical induction one can define conserved quantities on the horizon,
\begin{equation}
H^{Aretakis}_{L}[\Phi]=\left(\partial_{r}^{L+1}\Phi\right)|_{r=0},
\label{areG}
\end{equation}
which we call Aretakis quantities or constants. 
Like the full Aretakis charge (\ref{AREe}), they do not depend on the time $v$ on the horizon.
This result is generic for any extremal non-dilatonic black $p$-brane in (p+4) dimensions; even though depending only on the $AdS_{2}$ section of the whole scalar field, $\Phi(v,r)$, the influence of extra dimensions are encoded in the quantum number $L$. Due to the fact that extremal non-dilatonic black $p$-branes can be viewed as extremal dilatonic black holes in four dimensions as discussed in section 2, the conserved quantities found above hold in the near-horizon geometry of such black holes. So, the 
novel property in Eq.(\ref{areG}) is that it provides a higher-dimensional generalization for the Aretakis quantities. 

The near-horizon geometry of extreme RN black hole  corresponds to $p=0$ and the quantum numbers: $m=0$ and $l^{\prime}=0$; as a result $L$ reduces to $l$, and we recover  well-known results \cite{Lucietti:2012xr}. Generically, as said before
, the quantum number $L$ is a combination of the set of quantum numbers $\left\{m,l^{\prime},l\right\}$. As a consequence, for a particular choice of $L$ there is a degenerate spectrum of possible choices of $m,l^{\prime}$ and $l$ leading to same Aretakis quantities. The lower-order mode is covered by taking $L=p$; for this case 
the corresponding Aretakis constant is expressed in terms of the space-time dimension. 
For example, for $p=1$ (corresponding to extremal non-dilatonic black strings)  the lower-order conserved quantity at the horizon is given by the $2$-nd radial derivative of the scalar mode $\Phi(v,r)$
. In general, for arbitrary extremal non-dilatonic black $p$-brane, $\left(\partial_{r}^{p+1}\Phi\right)|_{r=0}$ are the lower-order conserved quantities on the horizon.       


\subsection{Late time tails}
\label{subsec3}

			In this section we will examine the physical consequences of the conserved Aretakis quantities and what happens at the late time regime. 
The first step is proceeding with a coordinate transformation on the $AdS_2$ section of the full near-horizon metric $AdS_{p+2}\times S^2$.  
We introduce the coordinates $(v,u)$, which are related to
 $(t,r)$ by: 
				\begin{equation}
				v=t-\frac{1}{r}, \,\,\, u=t+\frac{1}{r}.
				\end{equation}
The metric in this coordinates becomes:
\begin{equation}
ds^2_{AdS_{2}}=-\frac{4}{(u-v)^2}du dv.
\end{equation}
Since $-\infty<u,v<\infty$ it is convenient to compactify the metric by introducing new coordinates. 
By defining
\begin{equation}
U=\tan^{-1}u,\,\,\, V=\tan^{-1}v,
\end{equation} 
 the former metric takes the form
\begin{equation}
ds^2_{AdS_{2}}=-\frac{4}{\sin^2(U-V)}dU dV,
\end{equation}
where $U$ and $V$ 
are still outgoing and ingoing null coordinates, respectively. However, now
 they are restricted to the region $-\dfrac{\pi}{2}<V<U<\dfrac{\pi}{2}$. The future Poincar\'{e} horizon, previously located at $r=0$, becomes located at $U=\dfrac{\pi}{2}$ in the new coordinate frame. 

In terms of the new coordinates  
Eq.(\ref{kgvr}) becomes
\begin{equation}
\partial_{V}\partial_{U}\Phi(U,V)+\frac{L(L+1)}{\sin^2(U-V)}\Phi(U,V)=0.
\label{eqvu}
\end{equation}
Similarly to the radial equation (see Appendix \ref{appendix1}), the former equation can be solved in terms of hypergeometric functions. Nevertheless, for our purposes it is enough to take into account the asymptotic behavior near the horizon. Thus, we consider the asymptotic 
behavior near the horizon of the mode $\Phi(U,V)$ which corresponds to the limit $U\rightarrow\dfrac{\pi}{2}$ and $V\rightarrow\dfrac{\pi}{2}$:  
\begin{equation}
\Phi(U,V)\rightarrow (U-V)^a\Big(f(V)+\mathcal{O}(U-V)\Big).
\label{puv}
\end{equation}
Substituting this into Eq.(\ref{eqvu}) we found two roots for $a$ which are
\begin{equation}
a_{\pm}=\frac{1}{2}\pm \frac{1}{2}\sqrt{1+4\xi^2},
\end{equation}
or, in terms of $L$, 
\begin{equation}
a_{+}=L+1,\,\,\,a_{-}=L.
\end{equation}

Once again we will make the choice for normalizable boundary conditions 
so that only $a=a_{+}$ holds. 
 Now we are ready to evaluate the Aretakis quantities at 
 late times.  
 By substituting the asymptotic 
 behavior Eq.(\ref{puv}) in Eq.(\ref{eqvu}) and using the Leibniz rule
we 
find: 
\begin{equation}
H_{L}=\frac{(-1)^{L+1} (L+1)!}{2^{2(L+1)}}f(\frac{\pi}{2}).
\label{Hl}
\end{equation}
Note that our result recovered those ones found in \cite{Lucietti:2012xr} for $AdS_2$ space up to a normalization constant. For the lowest state, we have
\begin{equation}
H_{p}=\frac{(-1)^{p+1} (p+1)!}{2^{2(p+1)}}f\left(\frac{\pi}{2}\right),
\end{equation} 
which is completely expected since that has been describing a warped geometry of $AdS_2$. In particular, for $p=0$ the warped factor is trivial and then the lowest state for $AdS_2$ is recovered. Using Eq.(\ref{Hl}) one can write the asymptotic behaviour (\ref{puv}) near the horizon in terms of the constants:
\begin{equation}
\Phi(U,V)=\frac{(-4)^{L+1}}{(L+1)!}H_{L}(U-V)^{L+1}+...
\end{equation}
or, in terms of $(v,r)$ coordinates, we have
\begin{equation}
\Phi(v,r=0)=\frac{(-4)^{L+1}}{(L+1)!}\left(\frac{1}{v}\right)^{L+1}H_{L}+...
\end{equation}
where the ellipsis stand for subleading terms. Such result shows that $\Phi$ falls off as $v^{-L-1}$ at the late time while its $p$-th radial derivative remains constant along the horizon.  
Applying the operator $\partial_{r}^{L+1}$ in Eq.(\ref{kgvr}) in order to examine highest derivatives than 
the $(L+1)$-order at the horizon, we have
\begin{equation}
\begin{split}
0&=2\partial_{v}\partial_{r}^{L+2}\Phi(v,r)+\partial_{r}^{L+2}\left[r^2\partial_{r}\Phi(v,r)\right]-L(L+1) \partial_{r}^{L+1}\Phi(v,r)\\
&=2\partial_{v}\partial_{r}^{L+2}\Phi(v,r)-L(L+1) \partial_{r}^{L+1}\Phi(v,r)+\sum_{n=0}^{L+2}{L+2\choose {\it n}}\frac{2r^{2-n}}{(2-n)!}\partial_{r}^{(p+3-n)}\Phi(v,r),
\end{split}
\end{equation}
and evaluating at the horizon,
\begin{equation}
\begin{split}
0&=2\partial_{v}\partial_{r}^{L+2}\Phi(v,r)|_{r=0}+(L+2)(L+1)\partial_{r}^{L+1}\Phi(v,r)|_{r=0}-L(L+1)\partial_{r}^{L+1}\Phi(v,r)|_{r=0}\\
&=\partial_{v}\partial_{r}^{L+2}\Phi(v,r)|_{r=0}+(L+1)\partial_{r}^{L+1}\Phi(v,r)|_{r=0}.\\
\end{split}
\end{equation}
Through Eq.(\ref{areG}) we find
\begin{equation}
\partial_{r}^{L+2}\Phi(v,r)|_{r=0}=-(L+1)H_{L}v+g(r=0),
\end{equation}
where $g(r=0)$ is an arbitrary function evaluated at the horizon. The last equation shows that $\partial_{r}^{L+2}\Phi(v,r)|_{r=0}$ grows linearly with $v$ at late time. Thus it blows up as $v\rightarrow\infty$  suggesting the presence of instabilities for $H_{L}\neq 0$. 
By contrast if $H_{L}=0$, then $\partial_{r}^{L+2}\Phi(v,r)|_{r=0}$ is a constant and as long as this constant is different from zero $\partial_{r}^{L+3}\Phi(v,r)|_{r=0}$ blows up at late times.


\section{Conclusions}
\label{con}

In this work we have considered the near-horizon geometry of extremal non-dilatonic black $p$-branes in $D=p+4$ in the presence of a probe massless scalar field in order to explore the arising of conserved quantities on the horizon. First, we constructed the near-horizon metric in Gaussian null coordinates 
in order to define and compute  the Aretakis quantities on the horizon. We have argued that its near-horizon geometry becomes $AdS_{p+2}\times S^2$, 
and that  $AdS_{p+2}$ is written as a warped product between $AdS_2$ and $\mathcal{H}^p$ or, equivalently, a conformal compactification of $AdS_2\times S^{p}$. In addition, upon dimensional reduction by $p$ times, we have shown that the near-horizon geometry reduces to a warped geometry of $AdS_2\times S^2$. This result is a consequence of the fact that extremal dilatonic black holes can be viewed as extremal non-dilatonic black $p$-branes.

Finding the Aretakis quantities  required  knowledge of normalizable and non-normalizable modes, which were found by solving directly the wave equation.
These modes play an important role as the choice for normalizable modes by imposing boundary conditions leads to the vanishing of boundary contributions. In this respect, the Aretakis quantities depends only on the $AdS_2$ section of the scalar mode. Furthermore, we have obtained the explicit form for the Aretakis quantities and 
their relation to the $p$ spatial dimensions. For example, for the zero modes in the near-horizon geometry of RN black hole the first transverse derivative of $\Phi$ is conserved, whilst for extremal dilatonic black holes viewed as an extremal non-dilatonic black $p$-brane the $p$-order transverse derivative of $\Phi$ is conserved. 
Therefore the instability depends on the extra $p$ spatial dimensions.

The results obtained in this paper in some sense generalize
 those in the literature focused on extremal black holes \cite{Lucietti:2012xr}, but in contrast to the cases explored previously, we have explored extremal black $p$-branes whose horizon is not compact such that near-horizon theorems for black holes do not hold. It is noteworthy that in spite of this difference conserved Aretakis quantities can be defined. Further work will hopefully clarify the minimum conditions for a spacetime to exhibit these properties.

\appendix
\section{Explicit scalar modes in Gaussian null coordinates}
\label{appendix1}

In order to find regular solutions at the singular points,  after imposing boundary conditions it is convenient  to use the well-known Sturm-Liouville theory, see \cite{zettl2005sturm, boyce2017elementary}  for a detailed discussion. First of all, 
Eq. (\ref{KG1}) might be cast into the Sturm-Liouville form\footnote{We are following the same conventions of \cite{boyce2017elementary}.} 
by taking  the  transformation $z=\sin^{2}\alpha$ in Eq. (\ref{KG1}). Thus we have
\begin{equation}
L[\phi(z)]=\beta r(z)\phi(z),
\label{ST}
\end{equation}
where
\begin{eqnarray}
L[\phi(z)]&=&-\frac{d}{d z}\left[p(z)\frac{d \phi(z)}{d z}\right]+q(z)\phi(z),\\
p(z)&=&z^{p/2} (1-z)^{(1-p)/2},\\
q(z)&=&\frac{m^{2}_{0}}{4}\frac{z^{(p-2)/2}}{(1-z)^{(p+3)/2}}+\frac{l^{\prime}(l^{\prime}+p-2)}{4}\frac{z^{(p-4)/2}}{(1-z)^{(p+1)/2}},\\
r(z)&=&\frac{z^{(p-2)/2}}{(1-z)^{(p+1)/2}},\\
\beta&=&\frac{\xi^2}{4},
\end{eqnarray}  
valid  in the range $0<z<1$, and satisfying boundary conditions of the general form
\begin{equation}
\alpha_{1} \phi(0)+ \alpha_{2} \phi^{\prime}(0)=0,\quad \alpha_{3}\phi(1)+ \alpha_{4} \phi^{\prime}(1)=0,
\end{equation} 
where $\alpha$'s are constants. Such a setup  constitutes 
a well-posed Cauchy problem. 
 As we shall see later, the  equation above can be expressed in terms of hypergeometric equations, and as a consequence the solutions are obtained as power series around the singular points. Regularity conditions are required in order to get well-defined solutions at the singular points. 
 It is  convenient to first examine the behavior of the solutions of Eq. (\ref{ST}) near the origin $(z=0)$, and then to consider these solutions near the boundary, using the  linear transformations of hypergeometric functions which convert the argument of the function from $z$ to $(1-z)$. After that, quantization conditions must be fulfilled in order to have well-behaved solutions at the origin and the boundary. The solutions will be classified 
 as normalizable (square integrable) or non-normalizable (failing to be square integrable). 
 Let $\psi_1$ and $\psi_2$ be solutions of the Klein-Gordon equation, and their inner product be:
\begin{equation}
(\psi_1, \psi_2)=-i\int_{\Sigma_{v}} d\alpha d^{p+2}x \sqrt{-h}\, n^{\mu}\,(\psi_{1}^{\ast}\partial_{\mu}\psi_2-\psi_{2}\partial_{\mu}\psi_{1}^{\ast}), 
\label{norm}
\end{equation}
where  $\Sigma_{v}$ is a smooth space-like section, $h_{\mu\nu}$ is the induced metric on  $\Sigma_{v}$ and $n_{\alpha}$ is a unit normal vector to $\Sigma_{v}$. The case $\psi_1=\psi_2$ defines the norm of the solution;  if this integral converges, the solution is called normalizable (square integrable), while otherwise it is called non-normalizable.

In order to find the solutions 
we start by substituting the function
\begin{equation}
\phi(z)=(1-z)^{\lambda} z^{\beta} \varphi(z),
\end{equation}
into Eq.(\ref{ST}), and we have
\begin{eqnarray}
&& z(1-z)\frac{d^2 \varphi}{d z^2}+\left(\left(-2\lambda-2\beta-\frac{1}{2}\right)z+2\lambda+\frac{1}{2}p\right)\frac{d \varphi}{d z}+\nonumber\\
&+&
\left(-\left(\beta+\lambda\right)^2 +\frac{1}{2}\left(\lambda+\beta\right)+\frac{1}{4}\xi^2\right)\varphi=0,
\label{hyper}
\end{eqnarray}
which is a hypergeometric differential equation. The coefficients satisfy the following conditions:
\begin{eqnarray}
2\beta(2\beta+p-2)&=&l^{\prime}(l^{\prime}+p-2),\\
2\lambda(2\lambda-p-1)&=&m_{0}^2,
\end{eqnarray}
  the solutions of which  are given by:
\begin{eqnarray}
\beta&=&\frac{1}{2}l^{\prime},\, \frac{1}{2}\left(2-p-l^{\prime}\right),\\
\lambda&=&\lambda_{\pm}=\frac{1}{4}(p+1)\pm\frac{1}{4}\sqrt{(p+1)^2 +4m_{0}^2}.
\label{lambda1}
\end{eqnarray}

As described in \cite{Grad, abramowitz1965handbook}, the hypergeometric equation 
has two linear independent solutions which in turn  depend on $\beta$ and $\lambda$. 
For convenience and without loss of generality, we make the choice $\lambda=\lambda_{+}$. Furthermore, we must distinguish the cases according to the parity of the dimensional parameter $p$. 
 Hence the solutions near the origin, $z=0$, for $p$ odd, are:
\begin{eqnarray}
\label{var1}\phi_{1}(z)&=&z^{\frac{l^{\prime}}{2}}(1-z)^{\lambda_{+}}\, {_2}F_{1}(a,b;\frac{p}{2}+l^{\prime};z),\\
\label{var2}\phi_{2}(z)&=&z^{\frac{2-p-l^{\prime}}{2}}(1-z)^{\lambda_{+}} {_2}F_{1}\left(a-\frac{p}{2}-l^{\prime}+1,b-\frac{p}{2}-l^{\prime}+1;\frac{4-p-2l^{\prime}}{2};z\right),
\end{eqnarray}
where we have defined the following quantities:
\begin{eqnarray}
a&=&\lambda_{+}-\frac{1}{4}(1-\sqrt{1+4\xi^2})+\frac{l^{\prime}}{2},\\
b&=&\lambda_{+}-\frac{1}{4}(1+\sqrt{1+4\xi^2})+\frac{l^{\prime}}{2}.
\end{eqnarray}
Since we want  to obtain real solutions (\ref{var1},\ref{var2}),  the mass term should satisfy the following inequality,
\begin{equation}
m^{2}_{0}\geq-\frac{1}{4}(p+1)^2,
\end{equation}
which is the well-known Breitenlohner-Freedman bound \cite{BF1, BF2}.

Let us  consider 
first the case of even $p$, for which the first solution is well-defined at $z=0$. However, the second one is no longer well-defined at the origin. In such case, the second independent solution takes the form
\begin{equation}
\begin{split}
\tilde{\phi}_2(z)&= \phi_1(z)\ln(z)-\\
& -\bigg[\sum_{k=1}^{\frac{p-2}{2}+l^{\prime}}\frac{\left(\frac{p-2}{2}+l^{\prime}\right)!(k-1)!}{\left(\frac{p-2}{2}+l^{\prime}-k\right)!\left(1-a\right)_k \left(1-b\right)_k}\,(-z)^{-k}+\\
&+ \sum_{k=0}^\infty\frac{\left(a\right)_k \left(b\right)_k}{\left(\frac{p}{2}+l^{\prime}\right)_k\,k!}\,g_k\, z^k\bigg],
\end{split}
\end{equation}
where 
\begin{eqnarray}
g_k&=&\psi(a+k)+\psi(b+k)-\psi(1+k)-\psi(\frac{p}{2}+l^{\prime}+k),\\
\psi(z)&=&\frac{\Gamma^{\prime}(z)}{\Gamma(z)},\\
(a)_k&=&\frac{\Gamma(a+k)}{\Gamma(a)},
\label{pol}
\end{eqnarray}
are defined as 
in  \cite{DLMF}.

It is now necessary  to impose boundary conditions near the origin. Using the leading behavior of the solutions and the derivative of the hypergeometric functions \cite{abramowitz1965handbook}, 
we require that the boundary term disappears at the origin
. In other words, the second solution is unacceptable since the first term in Eq. (\ref{KGE}) vanishes for the classical solutions in the bulk and then the second term (boundary term) evaluated at the origin only vanishes for $\phi_1(z)$,
\begin{equation}
\int d^2 x\int d^{p+1}y \sqrt{-g}\,g^{zz}\partial_{z}\phi\Big|_{z=1}=0.
\label{I2}
\end{equation}
This property is reminiscent  of AdS/CFT  where for the contributions coming from the interior of the bulk the coupling with 
operators defined at the conformal boundary is not allowed.

The next step is to examine the behavior near the boundary, $z=1$ or $\alpha=\pi/2$. In order to do this, it is most convenient to  rewrite the argument of the hypergeometric function of the solution $\phi_1$ in terms of $z$. This is reached by using the linear transformations for hypergeometric functions prescribed in \cite{abramowitz1965handbook, DLMF}. 
In general, we can write down:
\begin{equation}
\phi_1 (z)=C \phi_3 (z)+ D \phi_4 (z),
\end{equation}
where $C$ and $D$ are constants while $\phi_3(z)$ and $\phi_4(z)$ represent the solutions near the boundary. Their explicit forms depend on $\nu=\dfrac{1}{2}\sqrt{(p+1)^2 +4 m^{2}_{0}}$, which is only the difference between $\lambda_{+}$ and $\lambda_{-}$.

\textbf{i)} The first case corresponds to $\nu$ being a non-integer number. Explicitly, we have
\begin{eqnarray}
\nonumber\phi_1 (z)&=&\frac{\Gamma(\frac{p}{2}+l^{\prime})\Gamma(\nu)}{\Gamma(e_{+})\Gamma(f_{+})}z^{\frac{l^{\prime}}{2}}(1-z)^{\lambda_{-}}{_2}F_{1}\left(e_{-},f_{-};(1-\nu);(1-z)\right)+\\
&+&\frac{\Gamma(\frac{p}{2}+l^{\prime})\Gamma(-\nu)}{\Gamma(e_{-})\Gamma(f_{-})}z^{\frac{l^{\prime}}{2}}(1-z)^{\lambda_{+}}{_2}F_{1}\left(e_{+},f_{+};(1+\nu);(1-z)\right).
\label{fc}
\end{eqnarray}
where we have defined the following quantities:
\begin{eqnarray}
e_{\pm}&=&\lambda_{\pm}-\frac{1}{4}(1+\sqrt{4\xi^2 +1})+\frac{l^{\prime}}{2},\\
f_{\pm}&=&\lambda_{\pm}-\frac{1}{4}(1-\sqrt{4\xi^2 +1})+\frac{l^{\prime}}{2},\\
C&=&\frac{\Gamma(\frac{p}{2}+l^{\prime})\Gamma(-\nu)}{\Gamma(e_{-})\Gamma(f_{-})},\label{EqCD}\\
D&=&\frac{\Gamma(\frac{p}{2}+l^{\prime})\Gamma(\nu)}{\Gamma(e_{+})\Gamma(f_{+})}\label{EqCD1}.
\end{eqnarray}

\textbf{ii)} The second one corresponds to $\nu$ being an integer number. We get
\begin{eqnarray}
\nonumber\phi_1 (z)&=&z^{\frac{l^{\prime}}{2}}(1-z)^{\lambda_{+}}\bigg\lbrace\frac{\Gamma(\frac{p}{2}+l^{\prime})\Gamma(\nu)}{\Gamma(a)\Gamma(b)}(1-z)^{-\nu}\sum_{k=0}^{\nu-1}\frac{\left(a-\nu\right)_{k} \left(b-\nu\right)_{k}}{k! \left(1-\nu\right)_k}(1-z)^k-\\
\nonumber&-&\frac{(-1)^{\nu}\Gamma\left(\frac{p}{2}+l^{\prime}\right)}{\Gamma(a-\nu)\Gamma(b-\nu)}\sum_{k=0}^{\infty}\frac{\left(a\right)_k \left(b\right)_k}{k!(k+\nu)!}(1-z)^{k}\big[\ln(1-z)-\psi(k+1)-\\
 &-&\psi(k+\nu+1)+\psi(a+k)+\psi(b+k)\big]
\bigg\rbrace,
\label{sc}
\end{eqnarray}
which is conveniently rewritten as follows
 \begin{eqnarray}
\nonumber\phi_1 (z)&=&z^{\frac{l^{\prime}}{2}}(1-z)^{\lambda_{-}}\frac{\Gamma(\frac{p}{2}+l^{\prime})\Gamma(\nu)}{\Gamma(a)\Gamma(b)}\sum_{k=0}^{\nu-1}\frac{\left(a-\nu\right)_{k} \left(b-\nu\right)_{k}}{k! \left(1-\nu\right)_k}(1-z)^k-\\
\nonumber&-&z^{\frac{l^{\prime}}{2}}(1-z)^{\lambda_{+}}\frac{(-1)^{\nu}\Gamma\left(\frac{p}{2}+l^{\prime}\right)}{\Gamma(a-\nu)\Gamma(b-\nu)}\sum_{k=0}^{\infty}\frac{\left(a\right)_k \left(b\right)_k}{k!(k+\nu)!}(1-z)^{k}\bigg[\ln(1-z)-\\
\nonumber&-&\psi(k+1)-\psi(k+\nu+1)+\psi\left(a+k\right)+\\
&+&\psi\left(b+k\right)\bigg],
\label{sc1}
\end{eqnarray}
where the constants in this case are given by:
\begin{eqnarray}
C&=&\frac{(-1)^{\nu}\Gamma(\frac{p}{2}+l^{\prime})}{\Gamma(a-\nu)\Gamma(b-\nu)},\label{eq11}\\
D&=&\frac{\Gamma(\frac{p}{2}+l^{\prime})\Gamma(\nu)}{\Gamma(a)\Gamma(b)}.\label{eq12}
\end{eqnarray}

\textbf{iii)}The last case is $\nu=0$. We have
\begin{eqnarray}
\nonumber\phi_1 (z)&=&z^{\frac{l^{\prime}}{2}}(1-z)^{\frac{1}{4}(p+1)}\bigg\lbrace\frac{\Gamma\left(\frac{p}{2}+l^{\prime}\right)}{\Gamma(a)\Gamma(b)}\sum_{k=0}^{\infty}\frac{\left(a\right)_k \left(b\right)_k}{(k!)^2}\bigg[2\psi(k+1)-\psi(a+k)-\psi(b+k)-\\
&-&\ln(1-z)\bigg](1-z)^k
\bigg\rbrace,
\end{eqnarray}
where the constants are
\begin{eqnarray}
C=D=\frac{\Gamma\left(\frac{p}{2}+l^{\prime}\right)}{\Gamma(a)\Gamma(b)}.
\end{eqnarray}

\section{Quantization conditions and (non)-normalizable modes}
\label{appendix2}

 Now, let us examine the regularity condition, first, for both cases \textbf{i)} and \textbf{ii)}. For these cases, it is clear that the leading terms of Eq.(\ref{fc},\ref{sc1}) fall off as $(1-z)^{\lambda_{\pm}}$ at the boundary, respectively. As a result, it  follows directly from Eq. (\ref{norm}) that for $\nu>1$ the solution $\phi_4(z)$ fails to be square integrable (normalizable) at $z=1$
; on the other hand, $\phi_3(z)$ is normalizable. So, as pointed out in \cite{Balasubramanian:1998sn}
, the combination of both modes sets up a non-normalizable mode. However, one can select the normalizable modes by eliminating the first term on the right hand side in Eqs.(\ref{fc},\ref{sc1})  by taking into consideration that $\Gamma$ functions in the denominator have poles when the following quantization condition is satisfied, namely,
\begin{equation}
\xi=\pm\sqrt{[2m+l^{\prime}+2\lambda_{+}][-1+2m+l^{\prime}+2\lambda_{+}]}. \quad m=0,1,2,...\,
\label{QC1}
\end{equation} 
Such a quantization condition leads to the vanishing of the coefficient $D$ and then the fluctuating modes are purely given by $\phi_{3}(z)$. Therefore, we conclude that the normalizable modes are selected by the above condition.  On the other hand, the pure non-normalizable modes are selected by the set of values
\begin{equation}
\xi=\pm\sqrt{[2m+l^{\prime}+2\lambda_{-}][-1+2m+l^{\prime}+2\lambda_{-}]}, \quad m=0,1,2,...\, 
\label{QC2}
\end{equation}
for the range of $\nu$ considered above
; in this case, the non-fluctuating modes are purely given by $\phi_4(z)$. 

A different scenario comes up for $0<\nu<1$. In this situation, both solutions are square integrable at the boundary for $\nu$ integer and non-integer, thus $\phi_1 (z)$ is normalizable.  The quantization conditions Eqs. (\ref{QC1}, \ref{QC2}) hold. However, in this case Eq.(\ref{QC1}) selects the $\phi_3 (z)$ normalizable mode whilst Eq. (\ref{QC2}) picks the $\phi_4 (z)$ normalizable mode. It is worth to point out that the leading term of $\phi_1 (z)$ falls off as $(1-z)^{\lambda_{-}}$ at the boundary since $(1-z)^{\lambda_{-}}$ dominates $(1-z)^{\lambda_{+}}$ near the boundary.

Lastly, in case $\textbf{iii)}$ (i.e.~when $\nu=0$), both $\phi_3 (z)$ and $\phi_4 (z)$ are square integrable.

\acknowledgments

We are grateful to Chris Pope for helpful discussions and collaboration on related topics. We would also like to thank Gonzalo Olmo and Albert Petrov for discussions. The work of M.C. is supported in part by the DOE (HEP) Award de-sc0013528, the Fay R. and Eugene L. Langberg Endowed Chair (M.C.) and the Slovenian Research Agency (ARRS No. P1-0306). P.J.P. would like to thank the Brazilian agency CAPES for the financial support (PDE/CAPES grant, process 88881.17175/2018-01) and Department of Physics and Astronomy, University of Pennsylvania, for the hospitality. 



\end{document}